\documentstyle[12pt]{article}
\begin{document}
\title{Stationary and Axisymmetric Perfect-Fluid Solutions with Conformal
Motion}

\author{Marc Mars$^\ast$ and 
Jos\'e M. M. Senovilla\thanks{Also at Laboratori de F\'{\i}sica Matem\`atica, 
IEC, Barcelona.} \\
Departament de F\'{\i}sica Fonamental, Universitat de Barcelona, \\
Diagonal 647, 08028 Barcelona, Spain.}
\maketitle
\begin{abstract}
Stationary and axisymmetric perfect-fluid metrics are studied under the
assumption of the existence of a conformal Killing vector field and in the
general case of differential rotation. The possible Lie algebras for the conformal
group and corresponding canonical line-elements are explicitly given. It turns out
that only four different cases appear, the abelian and other three called I, II and
III. We explicitly find all the solutions in the abelian and I cases. For the abelian
case the general solution depends on an arbitrary function of a single variable and
the perfect fluid satisfies the equation of state $\rho =p+$const. This class of metrics 
is the one presented recently by one of us. The general solution for case I is a
new Petrov type D metric, with the velocity vector outside the 2-space spanned by
the two principal null directions and a barotropic equation of state $\rho +3p=0$.
For the cases II and III, the general solution has been found {\it only} under 
the further assumption of a natural separation of variables Ansatz. The conformal
Killing vectors in the solutions that come out here are, in fact, homothetic. 
No barotropic equation of state exists in these metrics unless for a new 
Petrov type D solution belonging to case II and with $\rho +3p=0$ which
cannot be interpreted as an axially symmetric solution and such that the 
velocity vector points in the direction of one of the Killing vectors.
This solution has the previously unknown curious property that both
commuting Killing vectors are timelike everywhere.
\end{abstract}
\newpage
\section{Introduction}
This contribution deals with stationary and axisymmetric
differentially rotating perfect-fluid solutions of the Einstein field
equations admitting a conformal Killing vector field. In order to handle
the non-linear partial differential Einstein equations, several
special assumptions
are usually made to simplify the problem: Petrov type, irrotational fluids,
non-isometrical symmetries.
One of these mathematical simplifications is the conformal symmetry. 
Conformal symmetries play an important role in the case of perfect-fluid
solutions of the Einstein field equations in the stationary and axisymmetric
case. In fact, among the very few known exact solutions under such conditions,
two important families possess a conformal Killing vector: a family with rigid
rotation and Petrov type D, depending on three parameters \cite{S1}, and a
large family depending on an arbitrary function with differential rotation
and Petrov type D \cite{S2}, which includes the previous family
as its rigidly rotating limit, both found by one of us.

An attempt to classify all known exact solutions with a conformal symmetry has
been recently done by J. Castej\'on-Amenedo \& A.A. Coley \cite {CC} without 
restricting to any particular
matter contents of the space-time. The general case of perfect-fluid
stationary and axisymmetric exact solutions in the case of rigid rotation has
been considered in the last few years in some papers by D. Kramer
\cite{K1},\cite{K2} and D. Kramer \& J. Carot \cite{KC}. 
The first paper is restricted to the case in which
the conformal Killing vector commutes with both the two Killing vectors, or
equivalently, due to the orthogonal transitivity of the space-time, to the case
in which the conformal Killing is orthogonal to both the Killing vectors.   
It is found in this paper that
the only exact solutions under these assumptions are the Schwarzschild interior
solution which is static and conformally flat, a more symmetric
solution belonging to Herlt's class
and the general type D solution with the fluid vector lying in the two-plane
generated by the two repeated null directions of the Weyl tensor \cite{S1}.
The second paper by D. Kramer \cite{K2} considers the situation in which the
commutation of the conformal Killing vector with each of the Killing vectors is
an arbitrary linear combination of the Killings (without component in the
conformal Killing vector itself). The main result in this paper is that no
solution of the Einstein field
equations for a perfect-fluid energy-momentum tensor exists under the
assumptions above.
Finally, in the third mentioned paper by D. Kramer \& J. Carot \cite{KC} the
remaining case in which at least one of the Lie derivatives of the conformal
Killing vector along the Killing vectors has a non-vanishing component in
the conformal Killing vector itself is studied. The result is again that
no non-static solutions for perfect fluids exist under these hypotheses.

Our aim in this paper is to generalize these results on rigid rotation to the
more general case of differentially rotating perfect fluids by finding out
all the exact perfect-fluid solutions arising when a conformal motion is added
to the stationary and axial symmetry in the space-time. The amount of work that
represents to consider the fourteen inequivalent three-dimensional Lie algebras
arising when the orbits of one of the generators are closed is substantially
restricted due to a recent result by the authors \cite{MS} which states that
in an axially symmetric space-time (stationary or not) a conformal
Killing vector must necessarily commute with the axial Killing vector whenever
no more conformal symmetry exists in the space-time. This result is purely
geometric and does not depend on orthogonal transitivity or any matter contents
of the space-time. Using this result, it turns out that only four inequivalent
conformal Lie algebras are allowed, the abelian case and three other cases
called in this paper I, II and III.

The plan of the paper is as follows. In the second section the four different
allowed Lie algebras are established and the canonical forms of the metric, as
well as the explicit form of the conformal Killing vector in these coordinates
are given in each of the four cases. The Einstein field equations for
a perfect-fluid energy-momentum tensor involving only the components of
the Einstein tensor are also written down in this section.

In the third section the abelian case is exhausted. The main result
in this section is that the general solution of a differentially
rotating perfect fluid (including rigid rotation as a limit case)
with a conformal motion is, apart from the Schwarzschild
interior solution, the type D solution with an arbitrary function depending
on a single variable and equation of state $\rho =p+$const recently presented
by
one of us \cite{S2}. This solution was
found under completely different hypotheses involving mainly some conditions
on the Weyl tensor and it turned out to have a conformal motion.

The fourth section is devoted to the case I and again the general solution
in this case is found. Besides some static solutions we do not explicitly
consider and the Schwarzschild interior solution, the general solution in this
case is given by a new differentially rotating type D metric with the velocity
vector outside the 2-space spanned by the two principal null directions
and barotropic equation of state $\rho +3p=0$. Its rigid rotation limit is
static (as it must be because no solutions in the rigidly rotating case were
found in the papers mentioned above). 

In the fifth section the case II is considered. In this case, however, the
general solution is found only under the additional assumption of a separation
of variables Ansatz which is strongly indicated by the two previous sections.
The general solution under this condition is, in fact, homothetic and has
no barotropic equation of state unless for a new type D solution with
$\rho +3p=0$. This solution cannot be interpreted as an axially symmetric
space-time and the fluid velocity vector is proportional to one
of the Killing vectors (so that the solution is in some sense rigid)
but it is included here because it shows the interesting feature that both 
Killing vectors are timelike everywhere (obviously, at any point in the
space-time there exists a linear combination of them which is spacelike, but
this cannot be done globally). This solution was previously unknown, even though
it belongs to a ``rigid'' case which was extensively treated in \cite{K2}.

Finally, in the sixth section the remaining case III is treated. As in the
previous section, the general solution in this case is found only under the
separation of variables Ansatz. The general solution turns out to be again
homothetic and no particular case of them has a barotropic equation of state.

\section{Canonical Forms of the Metric}

A recent result due to the authors \cite{MS}
states that the mere existence of a regular symmetry axis for
the axial symmetry restricts severely the Bianchi type of the three-dimensional
Lie algebra generated by the Killing vector fields and the conformal Killing
vector field.  In fact, it is proven that the axial Killing vector
$\vec{\eta}$ must commute with both the timelike Killing $\vec{\xi}$ and the
conformal Killing vector $\vec{k}$. So we have necessarily
\begin{eqnarray*}
\left [ \vec{\xi},\vec{\eta} \right ] = \vec{0},  \hspace{2cm} \left[
\vec{k}, \vec{\eta} \right ] = \vec{0}, 
\end{eqnarray*}
while the commutation relation between the timelike Killing and the
conformal Killing is an arbitrary linear combination of $\vec{\xi},\vec{\eta}$
and $\vec{k}$ with constant coefficients.

It can be easily seen that only four non-isomorphic Lie algebras are possible
under these conditions. They can be written as
\begin{eqnarray}
\mbox{\bf Abelian Case} \hspace{3cm} & &
\left [\vec{\xi},\vec{k} \right ] =\vec{0},  \nonumber \\
\mbox{\bf Case I} \hspace{3cm} & &
\left [ \vec{\xi},\vec{k} \right ] = b \vec{k}, \label{Lie}  \\
\mbox{\bf Case II} \hspace{3cm} & &
\left [\vec{\xi},\vec{k} \right ] = b \vec{\xi}, \nonumber \\
\mbox{\bf Case III} \hspace{3cm} & &
\left[ \vec{\xi},\vec{k} \right ] = b \vec{\eta}, \nonumber
\end{eqnarray}
where $b$ is an arbitrary non-vanishing constant, which can still be set equal
to one. However, this constant has dimensions and we will not fix it to
any specified value.

Since the two Killing vector fields commute, there always exist coordinates
$t,\phi$ outside the axis of symmetry 
such that these vector fields are written as
$\vec{\xi}= \partial_t$ and $\vec{\eta}=\partial_{\phi}$. Moreover, we are
interested in non-convective rotating perfect fluids and then there also exist
coordinates $x$ and $y$ in which the metric line-element decomposes into two
orthogonal blocks (theorem of Papapetrou). For an account of the above
results and definitions, see for instance {\cite{Sn}}.
We can always diagonalise the $x,y$ block and write the metric 
in these coordinates as
\begin{eqnarray*}
ds^2= \frac{1}{\Psi^2(x,y)}\left[ -F(x,y)\left( \frac{}{}
dt+ P(x,y) d\phi \right )^2
+ \frac{Q^2(x,y)}{F(x,y)}d\phi^2+dx^2+dy^2 \right ], 
\end{eqnarray*}
where $F(x,y)$ must be a positive function.
Until now we have not restricted this metric to have a conformal symmetry.
Imposing the conformal Killing equations
\begin{eqnarray*}
{\cal{L}}_{\vec{k}}g_{\alpha\beta}=\nabla_{\alpha}k_{\beta}+\nabla_{\beta}
k_{\alpha}= 2\Phi g_{\alpha\beta}
\end{eqnarray*}
in the coordinates $\{t,\phi,x,y\}$, we can restrict the form of the
line-element
to the following four forms depending on the Bianchi type of the Lie
algebra.

\noindent {\bf Abelian Case}
\begin{eqnarray}
ds^2= \frac{1}{\Psi^2(x,y)}\left[ \frac{}{}-F(x)\left( \frac{}{} dt+ P(x) d\phi
\right )^2
+ \frac{Q^2(x)}{F(x)}d\phi^2+dx^2+dy^2 \right ], \label{metrab}
\end{eqnarray}
where $\Psi$ is the only function which depends on $y$. The conformal Killing
vector of this metric is given by
\begin{eqnarray*}
\vec{k}= \frac{\partial}{\partial y}.
\end{eqnarray*}

\noindent {\bf Case I}
\begin{eqnarray}
ds^2= \frac{1}{\Psi^2(x,y)}\left[ -b^2 M^2(y) dt^2+ L^2(x) \left( \frac{}{}
d\phi
+b N(y)dt \right)^2+dx^2+dy^2 \right ] , \label{metrI}
\end{eqnarray}
(see \cite{KC}) where the functions $M(y)$ and $N(y)$ are known because
they satisfy the following trivially integrable ordinary differential equations
\begin{eqnarray}
\dot{M}^2 = 1+\alpha  M^2, \hspace{2cm}
\dot{N} = \omega M, \label{MN}
\end{eqnarray}
where $\alpha$ and $\omega$ are arbitrary constants and the dot means derivative
with respect to the variable $y$. Therefore, in this case I, only
the two functions $\Psi$ and $L$ remain to be determined in order to specify 
the metric completely. The conformal Killing vector is
\begin{eqnarray*}
\vec{k}= e^{bt} \left ( - \frac{1}{b}\frac{\dot{M}}{M}
\frac{\partial}{\partial t} + \left (N \frac{\dot{M}}{M} 
- \omega M \right )
\frac{\partial}{\partial \phi} +  \frac{\partial}{\partial y} \right ).
\end{eqnarray*} 

\noindent {\bf Case II}
\begin{eqnarray}
ds^2= \frac{1}{\Psi^2(x,y)}\left[ -F(x)\left( \frac{}{} e^{-b y}
dt+ P(x) d\phi \right )^2+ \frac{Q^2(x)}{F(x)}d\phi^2+dx^2+dy^2 \right ].
\label{metrII}
\end{eqnarray}
So, in this case the dependence of the functions on the variable $y$ is
completely determined by the conformal Killing equations except for
the global conformal factor of the metric $\Psi(x,y)$. The conformal
Killing vector is in this case
\begin{eqnarray*}
\vec{k}= b t \frac{\partial}{\partial t} + \frac{\partial}{\partial y}.
\end{eqnarray*}

\noindent {\bf Case III}
\begin{eqnarray}
ds^2= \frac{1}{\Psi^2(x,y)}\left[ -F(x) dt^2
+ \frac{Q^2(x)}{F(x)} \left (\frac{}{}d\phi + \left(P(x)-b y\right ) dt \right 
)^2
+dx^2+dy^2 \right ], \label{metrIII}
\end{eqnarray}
where again, the explicit dependence on $y$ is explicitly
known except for the function $\Psi(x,y)$. The conformal Killing
vector reads now
\begin{eqnarray*}
\vec{k}= b t \frac{\partial}{\partial \phi} + \frac{\partial}{\partial y}.
\end{eqnarray*}
In each of the four cases, the relationship between the scale factor
of the conformal Killing equations $\Phi$, and the global conformal factor
of the metric $\Psi$, is given by
\begin{eqnarray*}
\Phi= -e^{a_1 t}\frac{\partial_{y} \Psi}{\Psi},
\end{eqnarray*}
where $a_1=b$ in the case I and vanishes for all the remaining cases.
As a consequence of this equation, the dependence of the function $\Psi(x,y)$
on the variable $y$ must be non-trivial because, otherwise, we would have
more proper isometries than the initially considered.

We are interested in non-convective perfect-fluid solutions of the
Einstein field equations in the general case of differential rotation.
The energy-momentum tensor is then
\begin{eqnarray*}
T_{\alpha\beta}= \left ( \rho + p \right ) u_\alpha u_\beta + p g_{\alpha\beta},
\end{eqnarray*}
where $\rho$ stands for the energy density, $p$ for the pressure of the fluid and the
fluid vector $\vec{u}$ lies in the two-plane generated by the two Killing
vector fields $\vec{\xi}$ and $\vec{\eta}$. 
In consequence, the fluid one-form is a linear combination of the
coordinate forms $\mbox{\boldmath$dt$}$ and $\mbox{\boldmath$d\phi$}$ at each
point of the space-time. 

We will write the Einstein field equations in orthonormal tetrads
$\{\mbox{\boldmath$\theta^{\alpha}$}\}$ chosen in
such a way that the fluid one-form $\mbox{\boldmath$u$}$ always lies on the
two-plane spanned by $\mbox{\boldmath$\theta^0$}$ and
$\mbox{\boldmath$\theta^1$}$ in each point. Therefore, we have 
\begin{eqnarray*}
\mbox{\boldmath$u$}=u_0 \mbox{\boldmath$\theta^0$}+ u_1 \mbox{\boldmath$
\theta^1$}, \hspace{15mm} u_{0}^2-u_{1}^2=1,
\end{eqnarray*}
and the Einstein field equations, in units where $c= 8 \pi G=1$, read
\begin{eqnarray*}
S_{00}  = \left(\rho + p\right ) u_0^2 - p, \hspace{15mm}\\
S_{01}  = \left ( \rho + p \right ) u_0 u_1, \hspace{16mm}\\
S_{11}  = \left ( \rho + p \right ) u_1^2 +p \hspace{15mm}\\
S_{22} = S_{33}  = p, \hspace{18mm}\\
S_{01}=S_{02}=S_{03} = S_{12}=S_{13}=S_{23} = 0,
\end{eqnarray*}
where $S_{\alpha\beta}$ stands for the Einstein tensor in the
$\{\mbox{\boldmath${\theta}^{\alpha}$}\}$ cobasis. 

These equations can be rewritten in terms of only the Einstein tensor,
while the calculation of $\rho,p$ and $\vec{u}$ is performed once
these equations are solved. The only non-trivially satisfied equations are
\begin{eqnarray}
\left ( S_{00}+S_{22} \right ) \left ( S_{11} - S_{22} \right ) -
S_{01}^2 = 0,\label{a} \\
S_{22} - S_{33} =  0, \label{b}\\
S_{23} = 0. \label{c}
\end{eqnarray}

\section{Abelian Case}

The orthonormal tetrad adapted to the form of the metric in this case
(\ref{metrab}) is given by
\begin{eqnarray*}
\mbox{\boldmath$\theta^0$}=\frac{1}{\Psi}\sqrt{F}\left(\frac{}{}
\mbox{\boldmath $dt$}+P\mbox{\boldmath$d\phi$} \right), \hspace{3mm}
\mbox{\boldmath$\theta^1$}=\frac{1}{\Psi}\frac{Q}{\sqrt{F}} \mbox{\boldmath
$d\phi$}, \hspace{3mm}
\mbox{\boldmath$\theta^2$}=\frac{1}{\Psi}\mbox{\boldmath$dx$}, \hspace{3mm}
\mbox{\boldmath$\theta^3$}= \frac{1}{\Psi}\mbox{\boldmath$dy$}.
\end{eqnarray*}
The equation $ S_{23}=0 $ reads
\begin{eqnarray*}
{\Psi}_{,xy} = 0,
\end{eqnarray*}
where the comma means partial derivative. This equation immediately gives
\begin{eqnarray*}
\Psi(x,y)= h(y) + g(x).
\end{eqnarray*}
The equation $S_{22}-S_{33}=0 $  takes the following form after dropping a
global factor $\Psi$
\begin{eqnarray*}
\left[ h(y)+g(x)\right] W(x) + 2  \left ( \ddot{h}(y) - g''(x)  \right ) = 0, 
\end{eqnarray*}
where $W(x)$ is an expression depending only on functions of $x$ (given below in
(\ref{G2233})) and the prime
denotes ordinary derivative with respect to the variable $x$.
Due to the fact that the function $h(y)$ cannot be a constant, it
follows from this equation the following three relations, which in particular
give us the explicit form of the functions $h(y)$ and $g(x)$
\begin{eqnarray}
\ddot{h}(y)  = \epsilon a^2 h(y) + c, \hspace{1.5cm}
g''(x) = - \epsilon a^2 g(x) + c , \label{gg} \\
W(x) \equiv - \frac{1}{2} \frac{F^2}{Q^2} {P'}^2 + \frac{Q''}{Q} - \frac{F'}{F}
\frac{Q'}{Q}+ \frac{1}{2}\frac{{F'}^2}{F^2}=-2 \epsilon a^2, \label{G2233}
\end{eqnarray}
where $\epsilon$ is a sign and $a$ and $c$ are arbitrary constants.
In the case that the constant $a$ is non-vanishing, the constant $c$
can be set equal to zero by adding to $g(x)$ and $h(y)$ appropriate constants. 
In fact, we have
\begin{eqnarray*}
\left( g - \frac{c}{\epsilon a^2} \right)'' = - \epsilon a^2 \left ( g - 
\frac{c}{\epsilon a^2} \right ), \hspace{1cm}
\left( h +\frac{c}{\epsilon a^2} \right)\mbox{\Large{$\ddot{}$}}=  
\epsilon a^2 \left ( h + \frac{c}{\epsilon a^2} \right ), 
\end{eqnarray*}
and therefore, renaming
\begin{eqnarray*}
g - \frac{c}{\epsilon a^2} \rightarrow g, \hspace{1.5cm}
h +\frac{c}{\epsilon a^2} \rightarrow h,
\end{eqnarray*}
we still have $\Psi(x,y)= g(x) + h(y)$ and the new functions $g$ and $h$ satisfy
(\ref{gg}) with $c=0$. The remaining Einstein equation (\ref{a}) takes the 
following form
\begin{eqnarray*}
\Sigma_1(x) \Psi^2(x,y) + \Sigma_2(x) \Psi(x,y) + \Sigma_3(x) =0, 
\end{eqnarray*}
where, as before, $\Sigma_{i}$ are expressions depending only on
functions of $x$. From this equation it follows that each $\Sigma_i$
must vanish. After some calculation involving the equation (\ref {G2233}), 
the resulting three equations can be rewritten as
\begin{eqnarray}
\left( \frac{FP''}{Q}-\frac{FP'Q'}{Q^2}+\frac{2F'P'}{Q}
\right)^2- \left(\frac{F''}{F}- \frac{F'Q'}{FQ}\right)\left( 8\epsilon a^2+
2\frac{Q''}{Q}+\frac{F''}{F}-\frac{F'Q'}{FQ}\right )  = 0,  \label{eqn1}  \\
-2 g''\left(\frac{Q''}{Q}+4\epsilon a^2\right ) + g'\left ( \frac{Q'''}{Q}
+\frac{Q''Q'}{Q^2}+8\epsilon a^2 \frac{Q'}{Q}\right )=0 ,
\hspace{20mm} \label{eqn2}  \\
-2 {g''}^2 +2 g'g''\frac{Q'}{Q}- {g'}^2\left( \frac{Q''}{Q}+2\epsilon a^2
\right ) =0 . \hspace{26mm} \label{eqn3}
\end{eqnarray}
At this point, we must distinguish between two cases depending on whether
$g'\neq0$ or not.

\vspace{3mm}

\noindent {\bf Case} $\mbox{\boldmath$ g' \neq 0$}$

\vspace{3mm}

The last two equations (\ref{eqn2}), (\ref{eqn3}) show that the function $Q(x)$ 
must satisfy two different differential equations. By adding the derivative
of equation (\ref{eqn3}) to (\ref{eqn2}) multiplied by $g'$ we find a 
differential relation for $Q(x)$ containing derivatives of this function up to 
the second order. We can now use equation (\ref{eqn3}) again in order to find an
expression containing only first derivatives of the function $Q(x)$. 
This expression is a second order polinomial in $Q'(x)$ and reads explicitly,
after factorization 
\begin{eqnarray*}
\left ( g' \frac{Q'}{Q} - 2 g'' \right ) \left ( g' g'' \frac{Q'}{Q} +
\epsilon a^2 {g'}^2- {g''}^2 \right ) =0.
\end{eqnarray*}
In consequence, two different possibilities arise from this equation, namely
\begin{eqnarray*}
&\mbox{a1)}& \hspace{2cm} g' \frac{Q'}{Q} -2 g'' =0\hspace{1cm}
\Longleftrightarrow  \hspace{1cm} Q=Q_0 {g'}^2 \\
&\mbox{a2)}& \hspace{2cm} g' g'' \frac{Q'}{Q} - {g''}^2 - \epsilon a^2
{g'}^2=0 \hspace{1cm}
\end{eqnarray*}
where $Q_0$ is a constant of integration.

\vspace{3mm}

\noindent {\bf Subcase} $\mbox{\boldmath $a1)$}$

\vspace{3mm}

Let us first analyse the subcase $a1)$ where $Q=Q_0{g'}^2$. It can be
trivially checked that both equations (\ref{eqn2}) (\ref{eqn3}) are
identically satisfied and there only remain the differential equations
(\ref{G2233}) and (\ref{eqn1}) to be
solved. From the explicit expression for $Q$ it follows that it satisfies
\begin{eqnarray*}
\frac{Q''}{Q}= \frac{1}{2} \frac{{Q'}^2}{Q^2} - 2\epsilon a^2.
\end{eqnarray*}
Using this expression, equation (\ref{G2233}) gives
\begin{eqnarray*}
{P'}^2\frac{F^2}{Q^2} = \left ( \frac{F'}{F} - \frac{Q'}{Q} \right )^2,
\end{eqnarray*}
which can be easily integrated to give $P= - \sigma \frac{Q}{F} + \nu$,
where $\sigma$ is a sign and $\nu$ is an integration constant. Substituting
$Q$ and $P$ into the remaining equation (\ref{eqn1}) gives $\left( 
\frac{{g''}^2}{{g'}^2} + \epsilon a^2 \right )^2=0$
from what it follows that $\epsilon=-1$ and $g''=ag'$. 
In the particular case when $a=0$ we have that $g'$ is a constant and in 
consequence
$c=0$. Redefining $Q_0 {g'}^2 \rightarrow Q_0$, the solution is given by
\begin{eqnarray*}
P=-\sigma\frac{Q}{F}+\nu, \hspace{1cm} g''=0, \hspace{1cm}
Q=Q_0, \hspace{1cm} \ddot{h}= 0,
\end{eqnarray*}
while when the constant $a$ is non vanishing we have that g satisfies
$g'=ag$ and therefore we have $Q=a^2 Q_0 g^2$. Redefining again $Q_0$ the
solution is
\begin{eqnarray}
P=-\sigma\frac{Q}{F}+\nu, \hspace{1cm} g'= a g, \hspace{1cm}
Q=Q_0 g^2, \hspace{1cm} \ddot{h}= - a^2 h^2 \label{sll}.
\end{eqnarray}
In both cases $F$ is an arbitrary function of the variable $x$. 
These solutions, however, are not perfect-fluid solutions of Einstein's
field equations because their Einstein's tensors do not have any
timelike eigenvector. This type of solutions can arise because in the quadratic
equation (\ref{a}) some non-perfect fluid solutions are included.

\vspace{3mm}

\noindent {\bf Subcase} $\mbox{\boldmath $a2)$}$

\vspace{3mm}

Let us now look at the subcase in which
\begin{eqnarray}
g' g'' \frac{Q'}{Q} - {g''}^2 + \epsilon a^2 {g'}^2 =0.\label{a2}
\end{eqnarray}
In this case, equation (\ref{eqn3}) gives after dropping a factor ${g'}^2$
\begin{eqnarray*}
\frac{Q''}{Q} + 4 \epsilon a^2 =0,
\end{eqnarray*}
which simplifies notably the equation (\ref{eqn1}) to give
\begin{eqnarray*}
\left( \frac{F^2P'}{Q} \right )' = \sigma \left ( F''- F' \frac{Q'}{Q}
\right ),
\end{eqnarray*}
where $\sigma$ is a sign.
Using this expression, the derivative of the expression (\ref{G2233})
takes the form 
\begin{eqnarray*}
\left (  F'' - F' \frac{Q'}{Q} \right ) \left ( \sigma \frac{F^2P'}{Q}
+ F \frac {Q'}{Q} - F' \right )=0. 
\end{eqnarray*}
It can be easily seen that the solution which is found when the second 
expression in round brackets vanishes
is exactly (\ref{sll}), so we need to consider only the other case in which
\begin{eqnarray*}
F'=\alpha Q, \hspace{1cm} \frac{F^2P'}{Q}=\beta
\end{eqnarray*}
where $\alpha$ and $\beta$ are constants. Equation (\ref{G2233}) becomes
\begin{eqnarray}
\beta^2- \alpha^2Q^2 + 2\alpha F Q' + 4\epsilon a^2F^2=0 \label{restr}
\end{eqnarray} 
and is now, despite its appearance, an algebraic relation between the
integration constants of the
equations. In order to satisfy the two remaining equations (\ref{eqn2}) and
(\ref{eqn3}),
we must distinguish between two cases depending on whether
$g''$ vanishes or not. If $g''=0$ we have that the constants $a$ and $c$
vanish. The condition (\ref{a2}) is identically satisfied, while equations
(\ref{eqn2}) and (\ref{eqn3}) give only $Q''=0$. The solution is therefore
\begin{eqnarray*}
g''= 0, \hspace{1cm} \ddot{h}=0,
\hspace{1cm} Q''=0, \hspace{1cm} F'=\alpha Q, \hspace{1cm} 
\frac{F^2P'}{Q}=\beta,
\end{eqnarray*}
with the constants restricted to satisfy the relation (\ref{restr}) with $a=0$.

When $g''$ is not vanishing, the condition $a2)$ can be integrated to give
\begin{eqnarray*}
Q=Q_0g'g'',
\end{eqnarray*}
where $Q_0$ is an integration constant. Using this expression
for $Q$ it is immediate to see that the
equations (\ref{eqn2}) (\ref{eqn3}) are identically satisfied and then
the solution is given by
\begin{eqnarray*}
g''= -\epsilon a^2 g, \hspace{1cm} \ddot{h}=\epsilon a^2 h,
\hspace{1cm} Q=Q_0 g' g'', \hspace{1cm} F'=\alpha Q, \hspace{1cm} 
\frac{F^2P'}{Q}=\beta,
\end{eqnarray*}
where the constants are restricted, as before, to satisfy the algebraic relation 
(\ref{restr}).

These two solutions are seen to be conformally flat and therefore, they both
are the Schwarzschild interior solution with constant density \cite{KSMH}.

\vspace{3mm}

\noindent {\bf Case} $\mbox{\boldmath$ g' = 0 $}$.

\vspace{3mm}

In the case when $g'=0$, the constant $g$ can be set equal to zero by
adding it to the function $h$. Thus, equations (\ref{eqn2}) and
(\ref{eqn3}) are trivially
satisfied and there only remain two differential equations to be satisfied by
the three functions $F,Q$ and $P$. In consequence we have a general family of
differentially rotating perfect-fluid solutions which depend on an arbitrary
function. Performing the change of functions and variables given by
\begin{eqnarray*}
F=m , \hspace{15mm} P=\frac{s}{m}, \hspace{15mm} Q^2= h m + s^2, \hspace{15mm}
dx=\frac{d\tilde{x}}{\sqrt{h m+ s^2}},
\end{eqnarray*}
the field equations (\ref{G2233}),(\ref{eqn1}) become
\begin{eqnarray*}
\ddot{s}^2+\ddot{h} \ddot{m}=0, \\
{\left ( h m+ s^2 \right )}\mbox{\Large{$\ddot{}$}} 
+ 4 \epsilon a^2 = \dot{s}^2+ \dot{h} \dot{m},
\end{eqnarray*}
where the dot means, only in these two equations, derivative with respect
to $\tilde{x}$. This family of solutions is a differentially rotating generalization
of a rigid solution due to one of us \cite{S1} and was found by the same author
\cite{S2} as the more general stationary and axisymmetric non-convective and
differentially rotating perfect-fluid solution satisfying the following
assumptions:

(i) Petrov type D,

(ii) The velocity vector lies in the two-plane spanned by the two repeated
principal null
directions of the Weyl tensor.

(iii) The Weyl tensor has vanishing magnetic part with respect to the fluid 
velocity vector.

This family of solutions has equation of state $p=\rho +$const. and belong to
the case D1DR in the classification scheme of \cite{Sn}. We have
therefore proven that this family of solutions, together with the Schwarzschild
interior solution are the only stationary axisymmetric non-convective
perfect-fluid solutions which admit a three-dimensional abelian conformal group with one
proper conformal Killing vector.

\section{Case I}

In this case the line-element can be cast into the form (\ref{metrI}) from which
the following orthonormal tetrad can be read
\begin{eqnarray*} 
\mbox{\boldmath$\theta^0$}=\frac{1}{\Psi} b M
\mbox{\boldmath $dt$}, \hspace{3mm}
\mbox{\boldmath$\theta^1$}=\frac{1}{\Psi}L \left ( \frac{}{}
\mbox{\boldmath$d\phi$} + b N \mbox{\boldmath$dt$} \right), \hspace{3mm}
\mbox{\boldmath$\theta^2$}=\frac{1}{\Psi}\mbox{\boldmath$dx$}, \hspace{3mm}
\mbox{\boldmath$\theta^3$}= \frac{1}{\Psi}\mbox{\boldmath$dy$}.
\end{eqnarray*}
Two of the Einstein field equations read in this case
\begin{eqnarray*}
S_{23}= 2 \Psi \Psi_{,xy}=0, \hspace{4cm} \\
S_{22}-S_{33}= -\Psi \left( \frac{L''}{L} \Psi + 2\Psi_{,yy} - 2 \Psi_{,xx}
-\alpha \Psi + \frac{1}{2} \omega^2 L^2 \Psi \right )=0.
\end{eqnarray*}
So, as in the previous case, the function $\Psi$ splits into the sum of a function 
of $x$ and a function of $y$
\begin{eqnarray*}
\Psi(x,y)=g(x)+ h(y),
\end{eqnarray*}
which are explicitly known because they are the
solutions of the ordinary 
differential equations
\begin{eqnarray*}
g''= -\epsilon a^2 g +c, \hspace{1.5cm} \ddot{h}=\epsilon a^2 h +c,
\end{eqnarray*}
while the function $L(x)$ satisfies the differential equation
\begin{eqnarray*}
\frac{L''}{L}-\alpha + 2\epsilon a^2 +\frac{1}{2} \omega^2 L^2 =0.
\end{eqnarray*}
Using these relations, the remaining field equation (\ref{a}) is
\begin{eqnarray}
\left( -2\frac{\dot{M}}{M}\dot{h}+2\ddot{h}\right ) \left (
2\frac{L'}{L}g'-2g''-\frac{1}{2}\omega^2 L^2 g- \frac{1}{2} \omega^2 L^2 h
\right ) - \dot{h}^2\omega^2 L^2 =0. \label{om}
\end{eqnarray}
In the case that the constant $\omega$ vanishes, the function $N$ is a constant
that can be set equal to zero by redefining the axial variable. In consequence,
it is obvious from the form of the metric that these solutions are static, and
we are not interested in them in this paper. Thus, we can restrict ourselves
to the case
$\omega \neq 0$. Equation (\ref{om}) can be seen completely equivalent
to the two following ordinary differential equations
\begin{eqnarray}
2\frac{L'}{L}g'-2g''-\frac{1}{2}\omega^2 L^2g - \frac{1}{2}
 \omega^2 \nu L^2 =0, \\
\left (-\frac{\dot{M}}{M}\dot{h}+ \ddot{h} \right ) \left( 
h-\nu \right) + \dot{h}^2=0 ,\label{y}
\end{eqnarray}
where $\nu$ is a constant of separation of variables. It is convenient
to define a new function $H(y)$ as
$ H\equiv h-\nu $,
which clearly satisfies the same differential equation as $h(y)$ with
the constant $c$ replaced by $c+\epsilon a^2 \nu$. This equation implies
\begin{eqnarray}
\dot{H}^2=\epsilon a^2 H^2 + 2\left ( \epsilon a^2 \nu + c \right ) H + H_0,
\label{H1}
\end{eqnarray}
where $H_0$ is a constant.
In terms of this function, equation (\ref{y}) can be integrated to give
\begin{eqnarray}
M=k_0H\dot{H} \label{M},
\end{eqnarray}
where $k_0$ is the constant of integration.

Using relation (\ref{H1}), it can be seen that the
differential equation $\dot{M}^2=1+\alpha M^2$ is satisfied by the function
(\ref{M}) if and only if
\begin{eqnarray*}
\alpha=4\epsilon a^2, \hspace{1cm} \epsilon a^2 \nu + c =0, \hspace{1cm}
k_0^2 H_{0}^2=1.
\end{eqnarray*}
Therefore, we can define a function $G(x)$ by $G\equiv g+ \nu,$
so that $\Psi=G+H$ and $G$ is the solution of the equation
\begin{eqnarray}
G''= - \epsilon a^2 G \label{G}.
\end{eqnarray}
The field equations are written in terms of $G$ as
\begin{eqnarray}
L'G'+ LG \left ( \epsilon a^2 - \frac{\omega^2}{4} L^2 \right )=0 , \label{L1}\\
L''+2L \left (\frac{\omega^2}{4} L^2 -\epsilon a^2 \right )=0.  \label{L2}
\end{eqnarray}
From these two equations it follows immediately the relation
\begin{eqnarray*}
2L'G'+G L''=0 \Longleftrightarrow \left ( L'G^2 \right )=0 \Longleftrightarrow 
L'G^2=\delta,
\end{eqnarray*}
where $\delta$ is constant. We must distinguish three different
possibilities.

First of all let us consider the case when the constant $\delta$ vanishes but 
$G\neq 0$, so that $L$ is a constant fixed by equations (\ref{L1}) and 
(\ref{L2}): $L^2=\frac{4\epsilon a^2}{\omega^2}$.
Consequently, we must have $\epsilon=1$ and the solution is given by
\begin{eqnarray*}
L=\frac{2a}{\omega}, \hspace{1cm} 
\dot{H}^2=a^2H^2+H_0, \hspace{1cm} 
{G'}^2=-a^2G^2+G_0, \hspace{1cm} M=\left |\frac{H \dot{H}}{H_0}\right |.
\end{eqnarray*}
Under these conditions the metric (\ref{metrI}) becomes conformally flat
and therefore this solution is again Schwarzschild interior.

In the second case, when $\delta \neq 0$, we have
$G= \sigma\sqrt{\frac{L'}{\delta}}$,
where $\sigma$ is a sign. Equation (\ref{G}) for $G$ reads now, in terms of
$L$
\begin{eqnarray*}
\frac{2L'''}{L'}- \frac{{L''}^2}{{L'}^2}+4\epsilon a^2=0,
\end{eqnarray*}
which can be seen incompatible with (\ref{L2}) unless $a=0$ and $\omega=0$,
against hipotheses. Thus, no solutions exist in this subcase.

It only remains the study of the third case, when $G=0$. In this situation, 
the function $L$ satisfies only the differential equation (\ref{L2})
which has a first integral
\begin{eqnarray*}
{L'}^2-2\epsilon a^2 L^2 + \frac{\omega^2}{4} L^4 - L_0 =0,
\end{eqnarray*}
where $L_0$ is a constant of integration. 
From (\ref{MN}) and (\ref{M}) we obtain the function $N$
\begin{eqnarray*}
N= \frac{1}{2}\omega k_0 H^2 + N_0,
\end{eqnarray*}
where $N_0$ is a constant of integration. This constant can be set equal to
zero redefining the axial coordinate by 
\begin{eqnarray}
\phi + b N_0 t \rightarrow \phi,\label{refi}
\end{eqnarray}
which does not change the form of the axial Killing vector field 
($\vec{\eta}= \partial_{\phi}$).
Recalling that $\frac{M^2}{H^2}= k_0^2 \dot{H}^2$, 
defining a new constant $\beta=bk_0$, 
redefining $\frac{\omega}{2}\rightarrow \omega$, renaming $L$ to $X$ 
and performing the change of variables
\begin{eqnarray*}
dx= \frac{dX}{\sqrt{2\epsilon a^2 X^2 - \omega^2 X^4 + L_0}},
\end{eqnarray*}
the metric (\ref{metrI}) is written finally as
\begin{eqnarray}
ds^2= -\beta^2 \dot{H}^2 dt^2 + X^2
\left(\frac{1}{H} d\phi+\beta\omega H dt\right)^2
+\frac{dX^2}{ H^2\left (2\epsilon a^2 X^2 - \omega^2 X^4 + L_0
\right ) } + \frac{dy^2 }{H^2} , \label{solu}
\end{eqnarray}
where $H(y)$ is explicitly known because is the solution of
\begin{eqnarray*}
\ddot{H}=\epsilon a^2 H.
\end{eqnarray*}
The fluid velocity vector is given by
\begin{eqnarray*}
\vec{u}=\frac{1}{\beta\sqrt{\dot{H}^2-H^2\omega^2X^2}}\left (
\frac{\partial}{\partial t} - 2 \beta \omega H^2 \frac{\partial}{\partial \phi}
\right )
\end{eqnarray*}
and therefore this solution is a differentially rotating perfect-fluid solution with the
rotation $\Omega$ given by
\begin{eqnarray*}
\Omega= -2\beta\omega H^2.
\end{eqnarray*}
The pressure and energy density of this solution are given by
\begin{eqnarray*}
p= \dot{H}^2 - H^2\omega^2 X^2, \hspace{11mm} 
\rho = -3 \left ( \dot{H}^2 - H^2 \omega^2 X^2 \right ),
\end{eqnarray*}
so that they are related by the equation of state $\rho+3p=0$.
Unluckily, no known matter is described by such an equation of state.

This solution is Petrov type D and its principal null directions are given by
\begin{eqnarray*}
{\mbox{\boldmath $l_1$}} = \beta \dot{H} {\mbox{\boldmath $dt$}} + \frac{1}{H}
{\mbox{\boldmath $dy$}}, \hspace{2cm} 
{\mbox{\boldmath $l_2$}} = \beta \dot{H} {\mbox{\boldmath $dt$}} - \frac{1}{H}
{\mbox{\boldmath $dy$}}.
\end{eqnarray*}
We see that the fluid velocity vector does not lie in the two-plane generated by
the two principal null directions and thus, this solution belongs to the class
D2DR in the classification scheme of \cite{Sn}. As far as we know, this is
the first solution found in that case. 
It has a regular symmetry axis provided that $L_0=1$ and its static limit can 
be trivially obtained by setting $\omega=0$. We conclude that this solution is, 
apart from static cases, the only stationary and axisymmetric non-convective 
perfect-fluid solution with a three-dimensional (proper) conformal group with 
Lie algebra given by case I in (\ref{Lie}).

\section{Case II}

To write the metric (\ref{metrII}) corresponding to this case 
we choose the orthonormal cobasis given by
\begin{eqnarray*}
\mbox{\boldmath$\theta^0$}=\frac{1}{\Psi}\sqrt{F}\left( e^{-by}
\mbox{\boldmath$dt$} +P \mbox{\boldmath $d\phi$}\right ), \hspace{3mm}
\mbox{\boldmath$\theta^1$}= \frac{1}{\Psi} \frac{Q}{\sqrt{F}} 
\mbox{\boldmath$d\phi$}, \hspace{3mm}
\mbox{\boldmath$\theta^2$}=\frac{1}{\Psi}\mbox{\boldmath$ dx$}, \hspace{3mm}
\mbox{\boldmath$\theta^3$}= \frac{1}{\Psi} \mbox{\boldmath $ dy$}.
\end{eqnarray*}
We will not consider the situation when the function $P(x)$ vanishes because this
corresponds to the well-known static case. 
The Einstein equations (\ref{c}) and (\ref{b}) are, respectively
\begin{eqnarray}
b\Psi \left ( \frac{1}{4}
\frac{P'}{P} \frac{P^2F^2}{Q^2} + \frac{1}{4}\frac{F'}{F} \right ) 
+ {\Psi}_{,xy}=0, \hspace{3.5cm} \label{G23??} \\
\Psi \left ( -\frac{1}{2}\frac{Q''}{Q} + \frac{1}{2}
\frac{Q'F'}{QF}
+\frac{1}{4} \frac{{P'}^2}{P^2} \frac{F^2P^2}{Q^2} - \frac{1}{4}
\frac{{F'}^2}{F^2}
 +\frac{b^2}{2} - \frac{b^2}{4} \frac{P^2F^2}{Q^2} \right ) + \Psi_{,xx}
-\Psi_{,yy} =0.\label{G2233??}
\end{eqnarray}
These two partial differential equations have the structure
\begin{eqnarray*}
\Psi_{,xy}=  H(x) \Psi, \hspace{2cm} \Psi_{,yy} - \Psi_{,xx} = G(x) \Psi,
\end{eqnarray*}
where $H(x)$ and $G(x)$ stand for the expressions in brackets in (\ref{G23??})
and (\ref{G2233??}) respectively. The next step now would be to find the general solution
of this pair of partial differential equations for the function $\Psi(x,y)$, but 
this is not a trivial task due mainly to the two facts that they are not ordinary but
partial differential equations and that $H(x)$ and $G(x)$ are not explicit functions of
$x$ but some differential relations between the unknowns $F(x)$, $P(x)$ and $Q(x)$. 
Therefore, in this paper we will
restrict ourselves to a particular solution of these equations. The Ansatz is suggested by
the previous cases and it consists in the separation of $\Psi(x,y)$ as a sum of a function
of $x$ and a function of $y$. That is, we will assume in this section the restrictive
Ansatz given by
\begin{eqnarray}
\Psi(x,y)= g(x) + h(y). \label{Ansat}
\end{eqnarray}
The first partial differential equation (\ref{G23??}) is now simply
\begin{eqnarray}
\frac{P'}{P} \frac{P^2F^2}{Q^2} + \frac{F'}{F} = 0 \label{eq0??}.
\end{eqnarray}
The second equation (\ref{G2233??}) implies on the one hand the usual relations which
determine the functions $g(x)$ and $h(y)$
\begin{eqnarray}
g''= -\epsilon a^2 g + c, \hspace{2cm} \ddot{h} = \epsilon a^2 + c \label {eqh},
\end{eqnarray}
and on the other hand the ordinary differential equation 
\begin{eqnarray}
-\frac{1}{2}\frac{Q''}{Q} + \frac{1}{2}
\frac{Q'F'}{QF}
+\frac{1}{4} \frac{{P'}^2}{P^2} \frac{F^2P^2}{Q^2} - \frac{1}{4}
\frac{{F'}^2}{F^2}
+\frac{b^2}{2} - \frac{b^2}{4} \frac{P^2F^2}{Q^2}  -\epsilon a^2 =0
\label{eq1??}.
\end{eqnarray}
Substituting the expression for $F'$ given in (\ref{eq0??})
into the relevant components of Einstein's tensor, we find
\begin{eqnarray*}
S_{01} &=& - \Psi \frac{PF}{Q} \left (  g'\frac{P'}{P} + \left(g+h \right)
Q(x)+ b
\dot{h} \right ),\\
S_{00}+S_{33}& = & \Psi \left( \frac{P^2F^2}{Q^2}\left [  g'\frac{P'}{P}
+ \left (g + h \right )R(x) \right]  + 2 \ddot{h} + 2 b \dot{h} \right ),\\
S_{11}-S_{22} & =& \Psi \left ( \frac{}{} A(x)g + B(x) + A(x) h \right ),
\end{eqnarray*}
where $R(x)$, $A(x)$ and $B(x)$ stand for the following expressions
\begin{eqnarray*}
R(x) & \equiv & \frac{1}{2} \frac{Q'P'}{QP}+ \frac{{P'}^2}{P^2} \frac{P^2F^2}{Q^2}
-\frac{1}{2}\frac{P''}{P}+ b^2, \\
A(x) & \equiv & \frac{P^2F^2}{Q^2} \left( \frac{3}{2} \frac{Q'P'}{QP} - \frac{1}{2}
\frac{P''}{P} + \frac{3}{2} \frac{{P'}^2}{P^2} \frac{P^2F^2}{Q^2} - \frac{1}{2}
\frac{{P'}^2}{P^2} + \frac{1}{2} b^2 \right ),
\\
B(x) & \equiv & 2g' \frac{Q'}{Q} + g' \frac{P'}{P} \frac{P^2F^2}{Q^2} - 2 g''.
\end{eqnarray*}
Then, the quadratic Einstein equation (\ref{a})
takes the following form, after dropping a global factor $\Psi$,
\begin{eqnarray}
\Sigma_1(x) h^2 + \Sigma_2(x) h + \Sigma_3(x) + \Sigma_4(x) h \dot{h} +
\Sigma_5(x)  \dot {h}=0 \label{Sigma??},
\end{eqnarray}
where we have used the relation
\begin{eqnarray*}
\dot{h}^2= \epsilon a^2 h^2 + 2 c h + h_0,
\end{eqnarray*}
which is a consequence of the differential equation (\ref{eqh}) for $h(y)$ 
and where $h_0$ is a constant.
It is not difficult to see that equation (\ref{Sigma??}) splits into the following three
subcases:

\noindent {\it  {\bf b1)}} When $\ddot{h}=c$ with $c$ non-vanishing or when $a \neq 0 $ and 
$h_0 \neq 0$.

\noindent
{\it {\bf b2)}} When $\dot{h}=\alpha$ is a constant.

\noindent
{\it {\bf b3)}} When $\dot{h}=ah$ with $a$ non-vanishing.
\vspace{3mm}

\noindent {\bf Subcase {\it {\bf b1)}}}

\vspace{3mm}

In this case equation (\ref{Sigma??}) can be seen equivalent to 
\begin{eqnarray*}
\Sigma_1(x)=0 ,\hspace{1cm} \Sigma_2(x)=0, \hspace{1cm} \Sigma_3(x)=0,
\hspace{1cm} \Sigma_4(x)=0, \hspace{1cm} \Sigma_5(x)=0. 
\end{eqnarray*}
The expression $\Sigma_5 - g \Sigma_4$ reads
\begin{equation}
g' \frac{Q'}{Q} - g'' = 0, \label{Qg}
\end{equation}
which is identically satisfied if the function $g$ is a constant. But we
know that, in this case, this constant could be absorved into the function $h(y)$, so that 
we would have $g =0$ and then also
$\Sigma_3(x) \equiv - \frac{P^2F^2}{Q^2} b^2 h_0 =0$.
But the constant $c$ must vanish due to the differential equation that the
function $g$ had to satisfy, so this equation would be incompatible with the hypothesis
that $h_0$ is non-vanishing. We must
therefore assume that $g' \neq 0 $. From (\ref{Qg}) immediately follows
$Q(x)= \beta g'$ where $\beta$ is a constant and, as a consequence, we have
$Q''=-\epsilon a^2 Q \label{Q''}$.
Using this formula, equation (\ref{eq1??}) together with the expression for
$\Sigma_4(x)$ gives 
$\frac{P^2F^2}{Q^2}= 1- \frac{\epsilon a^2}{b^2}$ from where it is very easy to see
that no solution of the system of differential equations exists in this subcase.

\hspace{3mm}

\noindent {\bf Subcase {\it {\bf b2)}}}

\hspace{3mm}

Here we have $\dot{h}= \alpha$ and 
equation (\ref{Sigma??}) splits into the three relations
\begin{eqnarray*}
\Sigma_1(x)=0, \hspace{1cm} \Sigma_2(x) + \alpha \Sigma_4(x)=0, \hspace{1cm}
\Sigma_3(x) + \alpha \Sigma_5(x)=0. 
\end{eqnarray*}
Using the fact that now the constants
$a$ and $c$ vanish it is easily found that these three equations are, respectively
\begin{eqnarray}
\Sigma_1=\frac{P^2F^2}{Q^2}R\left ( A - R \right ) =0, \hspace{3cm} \nonumber \\
 A \left ( \frac{P^2F^2}{Q^2}
\frac{P'}{P} g' + 2 b \alpha \right ) + \frac{P^2F^2}{Q^2} R \left (
B - 2 b \alpha - 2 \frac{P'}{P} g'\right ) =0, \nonumber \\
B \left ( \frac{P^2F^2}{Q^2} \frac{P'}{P} g' + 2 b \alpha \right )
- \frac{P^2F^2}{Q^2} \left (\frac{P'}{P} g' + b \alpha \right )^2 =0. \hspace{0.7cm} 
\label{termei}
\end{eqnarray}
From the form of $\Sigma_1$ we know that two different possibilities
arise depending on which factor vanishes. When the first factor vanishes
but the second is different from zero it is straightforward to see from the form of the
other two equations that 
\begin{eqnarray}
\frac{P^2F^2}{Q^2} = 2 \label{lmn2}
\end{eqnarray}
which simplifies greatly the system of partial differential equations and
allows us to see easily that no solutions exist. When the second factor
in $\Sigma_1$ vanishes but the first one is not zero, it can be deduced
the same relation (\ref{lmn2}) from the other two equations again. This fact
simplifies substantially the differential equations we are considering and
the general solution of Einstein field equations in this case is given by
\begin{eqnarray*}
P' = \delta P ,\hspace{1cm} Q' = - \delta Q ,\hspace{1cm} F =\sqrt{2}\frac{Q}{P},
\hspace{1cm} \dot{h}= \alpha ,\hspace{1cm} g'= - b\frac{\alpha}{\delta},
\end{eqnarray*}
where $\delta$ is an arbitrary non-vanishing constant. It can be seen that this solution 
possesses a four-dimensional group of symmetries acting multiply transitively on
three-dimensional timelike hypersurfaces and therefore they have much more symmetry
than the initially considered.

Finally, in order to exhaust this
subcase it remains to consider the possibility when both factors in $\Sigma_1$
vanish. In this case it cannot be found a relation of the type (\ref{lmn2})
and the situation is more complicated.
Therefore, in this case, the three functions $Q(x)$, $P(x)$ and $F(x)$ must
satisfy the five differential equations given by $R(x)=0$, $A(x)=0$, (\ref{eq0??}),
(\ref{eq1??}) and (\ref{termei}). It is much more difficult to see that
this system of differential equations is incompatible, but a rather long
calculation involving appropriate combinations of these equations and their
derivatives allows finally to prove that no solution exists in this subcase. 

\hspace{3mm}

\noindent {\bf Subcase {\it {\bf b3)}}}

\hspace{3mm}

In this subcase we have $\dot{h}=ah$ and equation
(\ref{Sigma??}) gives the three equations
\begin{eqnarray*}
\Sigma_1(x)+a\Sigma_4(x)=0, \hspace{1cm} \Sigma_2(x) + a \Sigma_5(x)=0, \hspace{1cm} \Sigma_3(x) =0,
\end{eqnarray*}
where now, taking into account that $h_0$ and $c$ vanish and $\epsilon=1$, we
have
\begin{eqnarray}
\Sigma_3=\frac{P^2F^2}{Q^2} \left ( Rg + \frac{P'}{P} g' \right )
\left ( Ag + B - Rg - \frac{P'}{P} g' \right )=0,\nonumber \\                                                            
\Sigma_2 + a \Sigma_5 = \left ( Ag + B - Rg - \frac{P'}{P} g' \right )
\left ( \frac{P^2F^2}{Q^2} R + 2a^2 + 2 ab \right ) + \label{esses} \\
+ \left ( Rg + \frac{P'}{P} g' \right )\left( \frac{P^2F^2}{Q^2} (A-R) -2 a b
\frac{P^2F^2}{Q^2}+ 2 a b + 2 a^2 \right )=0 , \nonumber \\
\Sigma_1 + a\Sigma_4= A \left ( \frac{P^2F^2}{Q^2}R + 2 a^2 + 2 a b \right )
- \frac{P^2F^2}{Q^2} \left ( R+ ab \right )^2 =0 . \nonumber
\end{eqnarray}
We now make an analysis similar to that made in the previous subcase. When
the first factor in $\Sigma_3$ vanishes but the second one is different from
zero, it can be seen from the other two equations (which take a much simpler
form) that, necessarily 
\begin{eqnarray}
\frac{P^2F^2}{Q^2} = 2 \frac{a+b}{b}. \label{lmnab}
\end{eqnarray}
This equation allows us to see after little effort that the system of five
differential equations for the three functions $P(x)$, $Q(x)$ and $F(x)$ 
is incompatible.

When the second factor in $\Sigma_3$ vanishes while the first
one is not zero, it can be found again from the other two equations the relation
(\ref{lmnab}). In this case, however, the system of equations has the general
solution 
\begin{eqnarray*}
\dot{h}=ah, \hspace{1cm} g''= - a^2 g ,\hspace{1cm} Q=Q_0, \hspace{1cm}
F=\frac{Q_0}{P}, \hspace{1cm} b=-2a ,
\end{eqnarray*}
where $Q_0$ is a non-vanishing arbitrary constant and $P(x)$ is an arbitrary
positive function of $x$. This is not, however, a perfect-fluid solution of
Einstein's field equations as can be seen from the fact that its Einstein tensor
does not have a timelike eigenvector.

Finally, we need to consider the situation when both factors in $\Sigma_3$
vanish. As in the previous subsection, this is the most difficult case because no
relation of type (\ref{lmnab}) can be extracted directly from the other two
equations. 
After a long calculation 
which involves the third relation in (\ref{esses}) as well as the equations (\ref{eq0??}), 
(\ref{eq1??}) together with the vanishing of both factors in brackets of $\Sigma_3$, it can 
be proven that there only exists solution when the function $g$ is a constant.
We can redefine $h$ with this constant and set $g=0$. Then, from expressions 
(\ref{esses}) we learn that $\Sigma_3$ and $\Sigma_2 + a \Sigma_5$ vanish identically. 
Einstein's field  equations for $Q$, $P$ and $F$ are now
\begin{eqnarray}
\frac{P'}{P} \frac{P^2F^2}{Q^2} + \frac{F'}{F} = 0 , \hspace{5cm} \nonumber \\
-\frac{1}{2}\frac{Q''}{Q} + \frac{1}{2} \frac{Q'F'}{QF}
+\frac{1}{4} \frac{{P'}^2}{P^2} \frac{F^2P^2}{Q^2} - \frac{1}{4} \frac{{F'}^2}{F^2}
+\frac{b^2}{2} - \frac{b^2}{4} \frac{P^2F^2}{Q^2}  - a^2 =0, \hspace{1cm} \label{solu??} \\ 
\left ( \frac{Q'F'}{QF} + \frac{{P'}^2}{P^2}\frac{P^2F^2}{Q^2}
+ \frac{F''}{F}
-\frac{{F'}^2}{F^2}
+ 4 a (a+b) + 2 b^2 \frac{P^2F^2}{Q^2}
 \right )\hspace{-1.5mm}\left ( \frac{F''}{F} - 
\frac{Q'F'}{QF} + b^2 \frac{P^2F^2}{Q^2} \right )  \nonumber \\
= \frac{P^2F^2}{Q^2}
\left ( \frac{P''}{P} + 2 \frac{P'F'}{PF} - \frac{Q'P'}{QP}- 2 b (a+b)
\right )^2. \hspace{2cm} \nonumber
\end{eqnarray}
The line-element is 
\begin{eqnarray}
ds^2= e^{-2ay} \left[ -F(x)\left( \frac{}{} e^{-b y}
dt+ P(x) d\phi \right )^2+ \frac{Q^2(x)}{F(x)}d\phi^2+dx^2+dy^2 \right ]
\label{metr2}
\end{eqnarray}
and the pressure and density are given by the expressions
\begin{eqnarray*}
p= e^{2ay} \left ( \frac{1}{2} \frac{Q'F'}{QF}+ \frac{1}{4} \frac{{P'}^2}{P^2}
\frac{P^2F^2}{Q^2} - \frac{1}{4}\frac{{F'}^2}{F^2} - \frac{b^2}{4}
\frac{P^2F^2}{Q^2} + a^2 + 2 a b + b^2 \right ), \\
\rho = e^{2ay} \left ( \frac{1}{2} \frac{Q'F'}{QF}+ \frac{1}{4} \frac{{P'}^2}{P^2}
\frac{P^2F^2}{Q^2} - \frac{1}{4}\frac{{F'}^2}{F^2} + \frac{3b^2}{4}
\frac{P^2F^2}{Q^2} + a^2 - b^2 \right ). \hspace{8mm}
\end{eqnarray*}
Now, the conformal Killing vector field becomes an homothetic vector
and therefore, in case an equation of state exists, it must be a linear relation
between the pressure and the density $p = \left ( \gamma + 1\right ) \rho$,
which in this case reads
\begin{eqnarray*}
\frac{1}{2} \frac{Q'F'}{QF}+ \frac{1}{4} \frac{{P'}^2}{P^2}
\frac{P^2F^2}{Q^2} - \frac{1}{4}\frac{{F'}^2}{F^2} + b^2
\frac{3\gamma + 4}{4\gamma}\frac{P^2F^2}{Q^2} + a^2 - b^2 -
\frac{2b}{\gamma} \left (a+b\right ) =0.
\end{eqnarray*}
It is not difficult to find the most general solution of (\ref{solu??}) satisfying this 
equation of state (and not leading to more symmetry in the spacetime), which is given by
\begin{eqnarray*}
P(x) = \frac{1}{\alpha} \frac{Q^2}{Q^2- \beta^2}, \hspace{2cm} 
F(x) = \frac{\alpha}{\beta} \left ( Q^2 - \beta^2 \right ), \hspace{2cm} b=-2a,
\end{eqnarray*}
where $\alpha$ and $\beta$ are arbitrary positive constants and the function
$Q(x)$ is restricted to satisfy the differential equation
\begin{eqnarray*}
\frac{Q''}{Q} + \frac{2a^2}{\beta^2} \left ( Q^2 - \beta^2 \right )=0.
\end{eqnarray*}
It can be easily checked that the Killing vector $\partial_{\phi}$, as well as $\partial_t$, is
timelike everywhere, and in fact, it can be seen that no linear combination of these two Killing
vectors vanishes at any point of the spacetime. Therefore, this solution is not 
axially symmetric. In consequence, the coordinate $\phi$ is not an angle variable and we will
replace it for $T$. Performing the change of variables given by
\begin{eqnarray*}
dx = \frac{dX}{\sqrt{{\delta}^2-  \frac{a^2 X^2}{\beta^2} \left ( X^2 - 2 \beta^2
\right )}}, \hspace{1cm} Y = \frac{1}{a} e^{-a y}
\end{eqnarray*}
where $\delta$ is an arbitary constant, the metric line-element takes the final form
\begin{eqnarray}
ds^2= \frac{\alpha\beta}{a^2 Y^2} dt^2 - \frac{X^2}{\alpha\beta} \left ( a Y dT
+ \frac{\alpha}{ a Y} dt
\right )^2 +  \frac{a^2 Y^2 dX^2}{{\delta}^2-  \frac{a^2 X^2}{\beta^2} \left ( X^2 - 2
\beta^2 \right )} + dY^2. \label{solTt}
\end{eqnarray}
Even though the two Killing vector fields of this solution, $\partial_T$ and 
$\partial_t$, are timelike everywhere, there obviously exists a linear combination of them
which is spacelike at any point of the spacetime . However, no Killing
vector in this solution is globally spacelike. In the study of the exact solutions
with a two-dimensional group of isometries acting on timelike surfaces, the
starting point is usually to assume a timelike and a spacelike Killing vector
(see {\cite{KSMH}}).
While this is obviously true locally, the previous solution shows that sometimes it
might be interesting to consider two different timelike Killing vectors form the very
beginning in order to find some of the exact solutions with a $G_2$ acting on $T_2$.
The fluid velocity vector of this solution is given by
\begin{eqnarray*}
\vec{u}= \frac{\beta}{\sqrt{X^2- \beta^2}} \frac{aY}{\sqrt{\alpha\beta}}
\frac{\partial}{\partial t}
\end{eqnarray*}
while the pressure and density are
\begin{eqnarray*}
p= \frac{1}{\beta^2 Y^2} \left ( \beta^2 - Q^2 \right ), \hspace{2cm}
\rho = - 3\frac{1}{\beta^2 Y^2} \left ( \beta^2 - Q^2 \right )
\end{eqnarray*}
so that the equation of state is again $\rho + 3 p =0$. Finally, this solution is Petrov type D with
the two repeated principal null directions given by
\begin{eqnarray*}
\vec{l}= - \frac{\sqrt{\alpha\beta}}{X} \frac{\partial}{\partial T} \pm \sqrt{ 
{\delta}^2-  \frac{a^2 X^2}{\beta^2} \left ( X^2 - 2 \beta^2 \right )} \frac{\partial}{\partial X},
\end{eqnarray*}
and therefore, the fluid velocity does not lie in the two-plane generated by the two principal
null directions. Although  Kramer \cite{K1} studied the general solution
of Einstein's field equations admitting two commuting Killing vectors and a
conformal Killing vector with the Lie algebra given in case II and with the
velocity vector proportional to one of the Killings, he
surprisingly did not find this solution we have just presented.

We have therefore found the most general solution of non-convective stationary and 
axisymmetric perfect-fluid metrics admitting a three-dimensional proper conformal group 
generated by a Lie algebra of type II in (\ref{Lie}) {\it only} for the case when the 
Ansatz of separation of variables (\ref{Ansat}) holds. 
The general case will be treated elsewhere.

\section{Case III}

All tensor quantities will be written throughout this section in the orthonormal
tetrad
\begin{eqnarray*}
\mbox{\boldmath$\theta^0$}=\frac{1}{\Psi}\sqrt{F} \mbox{\boldmath $dt$}, \hspace{3mm}
\mbox{\boldmath$\theta^1$}= \frac{1}{\Psi} \frac{Q}{\sqrt{F}} \left ( \frac{}{}
\mbox{\boldmath$d\phi$} + \left ( P- b y \right ) \mbox{\boldmath $dt$}
\right ), \hspace{3mm}
\mbox{\boldmath$\theta^2$}=\frac{1}{\Psi}
\mbox{\boldmath$ dx$}, \hspace{3mm}
\mbox{\boldmath$\theta^3$}= \frac{1}{\Psi} \mbox{\boldmath $ dy$},
\end{eqnarray*}
which is adapted to the line-element (\ref{metrIII}).

The Einstein equations (\ref{c}) and (\ref{b}) are, respectively
\begin{eqnarray*}
-\frac{b}{4}\Psi \frac{P'Q^2}{F^2} + {\Psi}_{,xy}=0, \hspace{3cm} \\
\Psi \left (\frac{1}{2} \frac{F'Q'}{FQ} - \frac{1}{4} \frac{{F'}^2}{F^2} - \frac{1}{2}
\frac{Q''}{Q} + \frac{1}{4}\frac{{P'}^2Q^2}{F^2} - \frac{b^2}{4}
\frac{Q^2}{F^2} \right )  + {\Psi}_{,xx} -{\Psi}_{,yy} =0.
\end{eqnarray*}
As in the previous case, these two partial differential equations have the structure
\begin{eqnarray*}
\Psi_{,xy}= H(x) \Psi, \hspace{2cm} \Psi_{,yy} - \Psi_{,xx} = G(x) \Psi,
\end{eqnarray*}
where again $H(x)$ and $G(x)$ stand for the expressions readable
from the equations above. In consequence, the considerations made
in the previous section hold in this case. We will not consider the general
case but we will assume the same Ansatz of the previous section: expression $H(x)$ vanishes.
The equations are now
\begin{eqnarray}
\Psi(x,y) = g(x) + h(y), \hspace{4mm} g'' = - \epsilon a^2 g + c, \hspace{4mm}
\ddot{h} = \epsilon a^2 h + c, \hspace{4mm} P' = 0, \nonumber \\
\frac{1}{2} \frac{F'Q'}{FQ} - \frac{1}{4} \frac{{F'}^2}{F^2} - \frac{1}{2}
\frac{Q''}{Q} - \frac{b^2}{4} \frac{Q^2}{F^2} - \epsilon a^2 = 0, \hspace{1cm} \label{ec}
\end{eqnarray}
where as usual $\epsilon$ is a sign and $a$ and $c$ are arbitrary constants.
The fact that the $P$ becomes a constant allows us to set $P=0$ by redefining the 
coordinate $\phi$ in a way analogous to (\ref{refi}).
With $P=0$ the relevant components of the Einstein tensor take the form
\begin{eqnarray}
\frac{1}{\Psi}\left (S_{00} + S_{33}\right) = \Psi \left ( \frac{1}{2}
\frac{F''}{F} + \frac{1}{2} \frac{F'Q'}{FQ} -\frac{1}{2} \frac{{F'}^2}{F^2}
+ 2 \epsilon a^2 \right ) + 2 g'' - g' {\frac{F'}{F}}, \label{S03}\\
\frac{1}{\Psi}\left (S_{11} - S_{22}\right) =  \Psi \left (
\frac{1}{2} \frac{F''}{F}
- \frac{1}{2}\frac{F'Q'}{FQ} 
- \frac{b^2}{2} \frac{Q^2}{F^2} \right )
- 2 g'' + \left ( 2 \frac{Q'}{Q} -\frac{F'}{F}\right ) g', \label{S12}\\
\frac{1}{\Psi}S_{01} = - b \frac{Q}{F}\dot{h}. \hspace{5cm} \nonumber
\end{eqnarray}
For further manipulation of these expressions we will put
\begin{eqnarray*}
\frac{1}{\Psi}\left (S_{00} + S_{33}\right) = Z_1(x) \Psi(x,y) + Z_2(x), \hspace{.7cm}
\frac{1}{\Psi}\left (S_{11} - S_{22}\right ) &=& V_1(x) \Psi(x,y) + V_2(x), 
\end{eqnarray*}
where the symbols $Z_1$, $Z_2$, $V_1$ and $V_2$ can be read explicitly
from expressions (\ref{S03}) and (\ref{S12}). The Einstein equation (\ref{a})
can be seen equivalent to the following three ordinary differential equations
\begin{eqnarray}
Z_1 V_1 = \epsilon a^2 b^2 \frac{Q^2}{F^2}, \hspace{2cm}
Z_1 V_2 + Z_2 V_1 = 2 b^2 \frac{Q^2}{F^2} \left ( c - \epsilon a^2 g \right ), \nonumber \\
Z_2 V_2 = b^2 \frac{Q^2}{F^2} \left ( \epsilon a^2 g^2 - 2 c g + h_0 \right ), 
\hspace{2cm} \label{eqc} 
\end{eqnarray}
where we have used the usual expression for $\dot{h}^2$, consequence of the
differential equation for $h(y)$.
We must now distinguish, as in previous sections, between two cases depending
on whether the function $g$ is a constant (which can be set equal to zero
by including it into $h$) or not. Let us begin by considering this last situation.

\hspace{3mm}

\noindent {\bf Case $\mbox{\boldmath$g' \neq 0 $}$}

\hspace{3mm}

We are now going to rewrite the system (\ref{eqc}) in a simpler form. In order
to do this we have to separate the cases when the constant $a$ vanishes or
not.

When $a \neq 0$ it follows from the first equation in the system that
$Z_1$ and $V_1$ are both different from zero. Dividing the second equation by
the first and taking into account that the constant $c$ can be set equal to
zero redefining the functions $g$ and $h$ by an additive constant (keeping
$g+h$ invariant), we find $\frac{V_2}{V_1} + \frac{Z_2}{Z_1} = - 2 g$. 
Dividing the third equation by the first one, we also have
$\frac{V_2}{V_1} \frac{Z_2}{Z_1} = g^2 + \frac{h_0}{\epsilon a^2}$.
Thus, $\frac{V_2}{V_1}$ and $\frac{Z_2}{Z_1}$ are the two solutions of the
quadratic algebraic equation $\left ( X+g \right )^2 = - \frac{\epsilon h_0}{a^2}$
for the unknown $X$, from where it follows that $-\epsilon h_0$ must be postive or zero. 
Defining a constant $n$ by means of $h_0 \equiv - \epsilon a^2 n^2$, we have found that 
equations (\ref{eqc}) are equivalent, in the case $a\neq 0$, to the system
\begin{eqnarray}
Z_1 V_1 = \epsilon a^2 b^2 \frac{Q^2}{F^2} ,\hspace{3mm}
Z_2 + \left ( g + n \right ) Z_1 =0 , \hspace{3mm} 
V_2 + \left ( g - n \right ) V_1 =0, \label{linel}
\end{eqnarray}
which is much simpler because two of the equations are linear in the second derivatives 
while in the previous system (\ref{eqc}) all equations
were quadratic. Then, we have in this case that the two functions $Q(x)$ and
$F(x)$ must satisfy the four differential equations given by (\ref{ec}) and
(\ref{linel}). It is not very difficult to see that this system of four
second order differential equations for two unkowns is incompatible. The proof
involves a very long calculation which combines the equations under
consideration together with their derivatives. The details of this calculation
will be omitted here.

When the constant $a$ vanishes, the system (\ref{eqc}) is even simpler
to handle because from the first equation it follows that either $Z_1=0$ or
$V_1=0$. Again, the two unknowns $Q(x)$ and $F(x)$ must
satisfy four ordinary differential equations. A simpler but still long
calculation shows that these four differential equations are incompatible and
that no solution for $Q(x)$ and $F(x)$ exists.

Therefore, we have that in the subcase ($\mbox{\boldmath$g'\neq0$}$)
no solutions of the Einstein's
field equations exist. We must then study the case when $g$ is a constant.

\hspace{3mm}

\noindent {\bf Case $\mbox{\boldmath $g'= 0$}$}

\hspace{3mm}

We already know that the constant $g$ can be made zero by adding it to
the function $h(y)$ and it follows from the equation for $g$ 
that the constant $c$ vanishes in this case. In consequence, the expressions
$Z_2(x)$ and $V_2(x)$ vanish identically and the equations (\ref{eqc}) 
(one of them is identically satisfied) take the form 
\begin{eqnarray*}
Z_1 V_1 = \epsilon a^2 b^2 \frac{Q^2}{F^2}, \hspace{1cm}
0= h_0 b^2 \frac{Q^2}{F^2} .
\end{eqnarray*}
Thus, the constant $h_0$ vanishes so that $\epsilon=1$ and the function
$h(y)$ is the solution of the differential equation
$\dot{h} = a h \Longrightarrow h=e^{ay}$.

The differential equations for the functions $Q(x)$ and $F(x)$ reduce simply to
\begin{eqnarray}
\frac{1}{2} \frac{F'Q'}{FQ} - \frac{1}{4} \frac{{F'}^2}{F^2} - \frac{1}{2}
\frac{Q''}{Q}  - \frac{b^2}{4} 
\frac{Q^2}{F^2} - a^2 = 0, \hspace{2.5cm} \nonumber\\
\left ( \frac{1}{2}
\frac{F''}{F} + \frac{1}{2} \frac{F'Q'}{FQ} -\frac{1}{2}
\frac{{F'}^2}{F^2}
+ 2 a^2 \right )
\left (\frac{1}{2} \frac{F''}{F}- \frac{1}{2}\frac{F'Q'}{FQ} 
- \frac{b^2}{2} \frac{Q^2}{F^2} \right )= a^2 b^2 \frac{Q^2}{F^2},
\label{solu???}
\end{eqnarray}
and the metric line-element is 
\begin{eqnarray}
ds^2= e^{-2ay} \left[ -F(x) dt^2 + \frac{Q^2(x)}{F(x)} \left (\frac{}{}d\phi 
-b y dt \right )^2 +dx^2+dy^2 \right ].
\label{metr3}
\end{eqnarray}
The pressure and energy density of the perfect fluid are
\begin{eqnarray*}
p= e^{2ay}\left ( \frac{1}{2} \frac{F'Q'}{FQ} - \frac{1}{4}\frac{{F'}^2}{F^2}
- \frac{b^2}{4} \frac{Q^2}{F^2} + a^2 \right ), \\
\rho = e^{2ay}\left (\frac{1}{2} \frac{F'Q'}{FQ} - \frac{1}{4}
\frac{{F'}^2}{F^2} +\frac{3b^2}{4} \frac{Q^2}{F^2} + a^2 \right ).
\end{eqnarray*}
Given that the conformal Killing vector is homothetic in this case, we know
that the equation of state, in case it exists, should be a linear relation 
$p = ( \gamma + 1 ) \rho$, where $\gamma$ is some constant. In our situation 
this linear equation of state reads
\begin{eqnarray*}
\frac{1}{2} \frac{F'Q'}{FQ} - \frac{1}{4}\frac{{F'}^2}{F^2}+
a^2 + \frac{3 \gamma + 4}{4\gamma} b^2 \frac{Q^2}{F^2} =0.
\end{eqnarray*}
It is not difficult to see that no solution of the system (\ref{solu???})
satisfies this last differential equation and therefore, unfortunately, the 
perfect-fluid solution given by the solution to (\ref{solu???}) does not have 
any equation of state.

The same considerations made at the end of the previous section hold in this case. We
have not found the most general stationary and axisymmetric solution for a non-convective
perfect-fluid source with a proper conformal Killing vector satisfying the Lie algebra
given by Case III in (\ref{Lie}), but we have proven that, under the assumption  
(\ref{Ansat}) of separation of variables, the general solution is given by (\ref{solu???}).
In this solution the conformal Killing is in fact homothetic and 
there never exists an equation of state.

To finish we will summarize the main results obtained in this paper in the
following table where the solutions of Einstein's field equations for a
stationary and axisymmetric space-time (or more generally a $G_2$ on $T_2$
spacetime) with a proper conformal motion and filled with a non-convective
perfect fluid are 
written down.

\vspace{1cm}

\begin{tabular}{|c||l@{}|}
\hline
Abelian Case & \begin{tabular}{@{}l|l@{}}
                ${\Psi}_{,x} \neq 0 $ \hspace{14.5mm} & Schwarzschild interior solution \\
                ${\Psi}_{,x} = 0$ & Solution in \cite{S2}
                \end{tabular}
\\
\hline
Case I & \begin{tabular}{@{}l|l@{}}
         ${\Psi}_{,x} \neq 0 $ \hspace{14.5mm} & Static solutions \\ 
         ${\Psi}_{,x} = 0$ & Solution (\ref{solu})  
         \end{tabular}
\\
\hline
Case II & \begin{tabular}{@{}l|l@{}}
          $\Psi$ non-separable &  ? \\
          \cline{2-2}
          $\Psi$ separable & \begin{tabular}{@{}l|l@{}}
                             ${\Psi}_{,x} \neq 0 $ & $G_4$ on $T_3$ solution \\ 
                             ${\Psi}_{,x} = 0$ & Solution (\ref{solu??})-(\ref{metr2}) including 
                               (\ref{solTt}) \hspace{3mm}
                             \end{tabular}
           \end{tabular}       
\\
\hline
Case III & \begin{tabular}{@{}l|l@{}}
            $\Psi$ non-separable &  ? \\ 
            \cline{2-2}
            $\Psi$ separable & \begin{tabular}{@{}l|l@{}}
		               ${\Psi}_{,x} \neq 0 $ & No solutions \hspace{39.4mm} \\ 
              
                               ${\Psi}_{,x} = 0$ & Solution (\ref{solu???})-(\ref{metr3})                                
                              \end{tabular} 
             \end{tabular}                 
\\
\hline
\end{tabular}

\vspace{1cm}

\section*{Acknowledgements}
M. Mars wishes to thank the {\it Direcci\'o General d'Universitats, Generalitat
de Catalunya}, for financial support.

All the tensors in this paper have been computed with the
algebraic computer programs CLASSI and REDUCE. All the very long calculations 
involved in the proofs of incompatibility of differential equations have
been performed with the essential help of REDUCE.

\end{document}